# On Einstein's Opponents, and Other Crackpots

Essay review of "Einsteins Gegner. Die öffentliche Kontroverse um die Relativitätstheorie in den 1920er Jahren" by Milena Wazeck.

*Jeroen van Dongen*

Institute for History and Foundations of Science, Utrecht University, the Netherlands

"This world is a strange madhouse. Currently, every coachman and every waiter is debating whether relativity theory is correct. Belief in this matter depends on political party affiliation."[1] Thus begins a letter by Albert Einstein to his one time close collaborator, mathematician Marcel Grossmann. It was written on 12 September 1920, just some three weeks after Berlin's Philharmonic Hall had hosted a rambunctious rally at which Einstein had been denounced as a fraud and scientific philistine.

The event, together with the public debate between Einstein and nobelist Philippe Lenard of a month later, constitutes the first apogee of, what has become known as, the anti-relativity movement. Sentiment against relativity had been brewing for some time, in various quarters; one of the speakers at the Berlin event, Ernst Gehrcke, spectroscopist and extra-ordinary professor of physics, had for instance started publishing against the theory already in 1911. However, since the very public announcement of the eclipse results of 1919, opposition against relativity had gained great momentum. These results had confirmed Einstein's predictions of light bending in the gravitational field of the Sun, and the ensuing publicity had propelled him into the international limelight; clearly, the enormous interest in relativity is manifest in the above letter to Grossmann.

The letter also attests to another aspect of the wider reception of relativity: belief in the theory would have been predisposed by political position. Indeed, Einstein was convinced that the fierceness of his opponents was foremost politically motivated;[2] after all, he was a prominent pacifist, democrat and Jew, hence an ideal scapegoat for German reactionaries, frustrated with the outcome of World War I and the November Revolution. The organizer of the event in the Berlin Philharmonic, Paul Weyland, has indeed been identified as a right-wing rabble-rouser with nationalist and *völkisch* ideals.[3] It thus seems obvious that the fiery character of the opposition to Einstein in the years of the Weimar Republic should be explained by the volatile nature of the latter's politics. Historians, in any case, have largely agreed with Einstein's assessment of his opponents' deeper motivations.[4]

---

[1] Einstein to Marcel Grossmann, 12 September 1920, Doc. 148 on pp. 428-430 in CPAE 10.
[2] (Einstein 1920).
[3] (Kleinert 1993), (Goenner 1993).
[4] See e.g. (Goenner 1993), (Hermann 1994), (Rowe 2006), (van Dongen 2007), and CPAE 7, pp. 101-113; CPAE 10, pp. xxxvii-xlii. Klaus Hentschel (1990) has however made a sharper distinction between politically and scientifically motivated opposition.

Milena Wazeck, in her recent book, "Einsteins Gegner",[5] has produced an extensive overview of the anti-relativity movement. She has focused foremost on the German opposition to Einstein, but, through its networking with American and other European anti-relativists, her story also provides insight into the broader international constitution of Einstein's opposition. Wazeck's analysis covers a wide period: starting with the earliest opposition, in the early 1910's, her study extends into the years of the Third Reich, when the movement for a while seemed to garner institutional recognition through its association with *Deutsche Physik*, with its ties to the Nazi ideology. Surprisingly, however, Wazeck finds that the fierceness of the opposition to Einstein was, in fact, *not* primarily due to the highly charged political atmosphere of the Weimar years.

Einstein's critics could disagree with relativity for a number of reasons: either they maintained a belief in the ether, or in the absolute nature of time, or e.g. found that the theory left too little room for various metaphysical perspectives. Wazeck's analysis shows that such positions could be found within the academic world, or beyond its perimeter, with "amateur" researchers, of which there were many in the first decades of the last century. These would consider themselves bona fide natural scientists, engaged in proper research in the tradition of the 19$^{th}$-century gentleman scientist, often believing that the perspective of their academic counterparts had become unduly compartmentalized. Wazeck further identifies three groups of opponents: physicists, philosophers, and, most interestingly, those that had found their own private solution to the riddles of the universe, based on their own newly found principles; in German, the "*Welträtsellöser*." But why did these intellectuals act so strongly against relativity, if not because of the heated political atmosphere of the Weimar years? Why, then, did they mount protest events, organize themselves in alternative academies and roll out newspaper campaigns?

It still seems clear that Weyland, at least, was primarily interested in garnering political capital. But, Wazeck argues, historians have had a biased perspective when looking at anti-relativists: they have studied the phenomenon foremost as viewed from the perspective of Einstein's biography, in which in particular the events of 1920 take a central role, and anti-relativists are seen as assailants. Too little have they asked the question what relativity theory did to these eventual opponents—one has overlooked the fact that they held that relativity had attacked *them*.

According to Wazeck, the successes of relativity marginalized those that had different ideas: academics of a different opinion, such as Gehrcke and Lenard, would increasingly find themselves on the margins of their profession, as their candidates for positions would be more easily overlooked and their institutional desires would be less likely to be fulfilled. *Welträtsellöser*, likewise, would find it ever more difficult to get their ideas published in respected journals, or be awarded speaking time at scientific meetings. Modern science, most prominently represented by Einstein's mathematical physics, had sidelined them. Thus, anti-relativists were only clamoring to the defense of their own ideas and scientific stature; to them, Einstein was the assailant.

Relativity threatened the "knowledge systems" of its opponents. These were quite varied, ranging from mechanical world pictures to occultist perspectives, but anti-relativists were quick to suspend their differences of opinion: facing a common enemy, they realized that they stood stronger, and their opposition would be more credible, when they stood together. Anti-relativists consequently built up networks to act against Einstein's theory in concert. This led to some success. For instance, the clamor about the theory in Germany contributed to the Nobel committee's delay in awarding its 1921 prize to Einstein, and to the particular choice of subject for which he finally did receive it: his account of the photo-electric effect, instead of the controversial theory of relativity.[6]

---

[5] (Wazeck 2009).

One of the most intriguing episodes that Wazeck has unearthed is the creation of the "Academy of Nations", in 1921:[7] the networks of Einstein opponents around Gehrcke in Germany and Arvid Reutherdahl, engineer and dean at the College of St. Thomas in St. Paul, Minnesota, organized themselves in an Academy that would bestow honorary degrees and award prizes. It consisted of an international board, and local, national chapters. The Academy of Nations represented the cream of the crop of anti-relativists (along with, of course, Lenard, who was not a member, and seemed to wish to retain some distance to the effort), and being invited to its board was considered a great honor. The Academy's charter was to counter the fragmentation of academic scholarship, and to promote a "true", "good" and "free" science—it aspired to act as an arbitrating international "*Wahrheitsbund*" ["truth society"]. First, however, relativity had to be fought back; an underlying motivation for the Academy's foundation had of course been to give Einstein's opponents more institutional credibility. The suggestion was that one was not dealing with a marginalized movement at all, but a prominent and thriving international association.

Nevertheless, anti-relativists were convinced that their opinions were being suppressed. Indeed, many believed that conspiracies were at work that thwarted the promotion of their ideas. The fact that for them relativity was obviously wrong, yet still so very successful, strengthened the contention that a plot was at play—and some anti-relativists were convinced that the co-conspirators were Jewish. Jews were held to dominate both the newspaper business and the new discipline of theoretical physics; they could thus easily advertize one of their own (Einstein) and his fallacious work (relativity). Gehrcke, for instance, kept emphasizing that the successes of relativity could only be explained by a state of "mass hypnosis", brought about by excessive and one-sided reporting.[8] Such a qualification resonated with familiar anti-Semitic reasonings (it was a known anti-Semite strategy to claim Jewish hyping), and was well received in right-wing media. Yet, Wazeck denies any overt anti-Semitic motivations on Gehrcke's part; in her perspective, a crucial distinction seems to be that he did not necessarily primarily intend to promote the rightist cause, as Weyland appeared to have attempted. These two then rather become each other's mirror image: Weyland co-opted the anti-relativistic stance to further his larger political desires, while Gehrcke echoed rightist argumentations, as these were certain to produce a welcome audience for his struggle against relativity.

Wazeck, however, does not deny that Gehrcke et al. did take a political stance while acting against relativity. She argues that they did not do so directly, simply by criticizing Einstein, but rather because of the fact that they did not hold back their criticism in a context in which this could only have obvious political connotations. In 1922, for example, Einstein cancelled his plenary lecture at the centennial meeting of the Society of German Natural Scientists and Physicians as he, and apparently the authorities too, feared violence against his person; only a few months before his friend Walther Rathenau, Foreign Minister and another prominent Jewish liberal, had been assassinated, and one feared a similar fate for Einstein. Anti-relativists did not show any moderation and went ahead with their planned protests at the event, playing down the threat and seeing an opportunity to promote their cause in the absence of the great man himself. Such a course of action one could find distasteful, but anti-relativists maintained that they were oblivious to any politicking, and were really only interested in matter-of-fact discussions on the merits of relativity theory. Wazeck justly points out that by not taking a position against the threat of violence anti-relativists did, in fact, take a political stance; being "*unpolitisch*" was simply not an actual option. Nevertheless, the claim that one was really only interested in factual debate strongly rings dishonest when riding a wave of rightist violence, particularly also in light of the anti-Semitic resonances indicated above. Wazeck may take the anti-relativists perhaps a bit too easily at their word when she follows their comfortable self-assessment—apolitical—and exhibits reluctance to consider a warmer coming together of, at least for some, political preference and anti-relativism.

---

[6] See (Friedman 2001), pp. 119-140.
[7] See also (Wazeck 2005a).
[8] See e.g. his collection (Gehrcke 1924).

Conspiracy theories tend to do well in uncertain times: they create order in chaos. Hence, they thrived in post-World War I Germany. Just as there is no real point in debating conspiracy theorists, there was no point in explaining relativity to anti-relativists, Wazeck astutely observes. Their strong opposition was not due to a lack of understanding, but rather the reaction to a perceived threat. Furthermore, anti-relativists were convinced of their own ideas, and were really only interested in pushing through their own theories; any explanation of relativity would not likely have changed their minds. Initially, relativists, and in particular Einstein himself, were willing to engage in correspondence or debate with their critics. By the early 1920's, however, they concluded that sufficient common ground was lacking, and likely chose not to further waste any valuable time.

The *Welträtsellöser* would thus, eventually, be marginalized in the most effective fashion: they would simply be ignored. Of course, this solicited the loudest possible protests. Such a reaction reflects that of many others whose ideas have been deemed out of order by scientific officialdom, and the dynamic outlined by Wazeck should be quite familiar to today's physicists. Indeed, the actions of many of Einstein's opponents resemble those of the thinkers now often referred to as, in perhaps an all too derisive manner, "crackpots". It thus appears that this phenomenon is at least as old as the existence of institutionalized science, which arbitrates authoritatively what is, and what is not, sound scientific practice and established truth; crackpots, with their own unshakable beliefs, in the end rather deny that authority than give up their ideas.

It has long been clear that dismissing the anti-relativists' objections as those of an assortment of dimwits who simply did not get it, as physicists intuitively have tended to do, does not suffice. Wazeck has undoubtedly brought the discussion of anti-relativists to yet a higher level by looking beyond the conditioning of Einstein's opponents by the national political arena of the Weimar years. In her analysis, as said, their fierceness was the consequence of the uncertainty felt when they were confronted with a new "knowledge system", in conflict with theirs, while at the same time the power to decide what constituted good and valuable science shifted away from them: their resistance to relativity was so intense and, often, malignant because they were struggling against their marginalization. This is a compelling analysis, even if, in the opinion of this reviewer, the conservative convictions of many anti-relativists dovetailed too comfortably with their opposition to Einstein to relegate these entirely to a secondary role.

Wazeck's book deserves further praise for the enormous wealth of archival material it discusses for the first time: she has worked through the Gehrcke and Reuterdahl archives, and has been closely involved in making parts of the first available on the internet.[9] All this deserves praise, as these efforts stand to make lasting contributions to the study of the history of relativity. Yet, this still does not mean that the study of anti-relativists has been exhausted. For instance, there is little discussion of the French reception of relativity in Wazeck's book, which might however provide an interesting counterpoint to her American and German stories (the French seem to have remained largely outside of the Gehrcke and Reutherdahl networks).[10] In any case, Wazeck has analyzed not only much new and interesting source material, but also presented a valuable and often convincing analysis of the anti-relativity movement. Hopefully, she will soon present her work to the scholarly community in English herself, by contributing an extensive review paper, or, even better, a translation of her wonderful book.

---

[9] The website is: http://echo.mpiwg-berlin.mpg.de/content/modernphysics/gehrcke. For more on the Gehrcke archive, see also (Wazeck 2005b).
[10] On the French reception of relativity, see e.g. (Paty 1987), (Biezunski 1991).